\documentclass[12pt]{article}

\input{tcilatex}

\begin{document}

\bigskip \baselineskip0.8cm \textwidth16.5 cm

\begin{center}
\textbf{Critical Exponents of the 3-dimensional Blume-Capel model on a
cellular automaton}

A. \"{O}zkan$^{\ast }$, N. Sefero\u{g}lu$^{\ast }$ and B. Kutlu$^{\ast \ast
} $

$^{\ast }$ Gazi \"{U}niversitesi, Fen Bilimleri Enstit\"{u}s\"{u}, Fizik
Anabilim Dal\i , Ankara, Turkey

$^{\ast \ast }$ Gazi \"{U}niversitesi, Fen -Edebiyat Fak\"{u}ltesi, Fizik B%
\"{o}l\"{u}m\"{u}, 06500 Teknikokullar, Ankara, Turkey

e-mail: bkutlu@gazi.edu.tr
\end{center}

\textbf{Abstract}

{\small The static critical exponents of the three dimensional Blume-Capel
model which has a tricritical point at }$D/J=2.82${\small \ value are
estimated for the standard and the cooling algorithms which improved from
Creutz Cellular Automaton. The analysis of the data using the finite-size
scaling and power law relations reproduce their well-established values in
the }$D/J<3${\small \ and }$D/J<2.8${\small \ parameter region at standard
and cooling algorithm, respectively. For the cooling algorithm at }$D/J=2.8$%
{\small \ value of single-ion anisotropy parameter , the static critical
exponents are estimated as }$\beta =0.31${\small , }$\gamma =\gamma ^{\prime
}=1.6${\small , }$\alpha =\alpha ^{\prime }=0.32${\small \ and }$\nu =0.87$%
{\small \ . These values are different from }$\beta =0.31${\small , }$\gamma
=\gamma ^{\prime }=1.25${\small , }$\alpha =\alpha ^{\prime }=0.12${\small \
and }$\nu =0.64${\small \ universal values. This case indicated that the BC
model exhibit an ununiversal critical behavior at the }$D/J=2.8${\small \
parameter value near the tricrital point( }$D/J=2.82${\small ). The
simulations carried out on a simple cubic lattice with periodic boundary
conditions.}


\textbf{1 . Introduction}

One of the interesting problems in the study of phase transitions of model
systems is to determine the order of phase transition and universality of
the systems. Blume-Capel model which exhibits a tricritical behavior has
been of considerable interest in recent years. The Hamiltonian of the model
is given by,%
\begin{equation}
H_{I}=-J\sum_{<ij>}S_{i}S_{j}+D\sum_{i}S_{i}^{2}
\end{equation}%
where $s_{i}=-1,0,1$ and the first sum is carried out over all
nearest-neighboring (nn) spin pairs on a three-dimensional simple cubic
lattice. The parameters of the $J$ and $D$ are the bilinear interaction
energy and single-ion anisotropy constant, respectively. The BC\ model has
been applied with success in many different physical situations such as the
magnetic phase transitions$[1,2]$, structural transitions$[3]$, dilute Ising
ferromagnets$[4]$ and staged intercalation compounds$[5]$. It is not exactly
solvable in three-dimension, but it has been studied over finite and
infinite d-dimensional lattices by means of many different simulation and
approximate techniques. An extensive analysis of this model for
three-dimensional lattices was made using the mean-field approximation$[1,2]$%
, the effective field theory$[6,7]$, the Bethe-Peierls approximation$[8,9]$,
the series expansion methods$[10,11]$, the self-consistent Ornsteien Zernike
(SCOZ) approximation $[12,13]$, \ the renormalization group theory$[14,15]$,
the cluster variation method$[16,17]$, the Monte Carlo method $[18-22]$,
cellular automaton$[23]$. Most of these analysis predict in the BC model the
existence of a tricritical point at which the phase transition changes from
second-order to first order for a $D/J$ value in the interval $2.7<\ D/J<2.9$
on the simple cubic lattice. The problem of identifying the tricritical
point is particularly difficult in numerical simulations. Difficulties have
arisen in the inaccessible of the entire region of metastable and unstable
states properties of first order phase transitions.

In previous paper$\left[ 23\right] $, we investigated the tricritical
behavior of the 3-d Blume-Capel model using an improved algorithm from
Creutz Cellular automaton (CCA) Ising model on a simple cubic lattice. The
phase diagram characterizing phase transition and the tricritical point
value of the model is obtained. For determining of the tricritical point,
the thermodynamic quantities are computed using two different procedures
which called as the standard and the cooling algorithm for the anisotropy
parameter values in the interval $3\geq D/J\geq -8$. The simulations confirm
the existence of a tricritical point at which the phase transition changes
from second-order to first order at the $D/J$ $=2.82$ value for the cooling
algorithm. The simulations indicates that the cooling algorithm is a
suitable procedure for the calculations near the first order phase
transition region, and the cooling rate is an important parameter in the
determining of the phase boundary. The CCA\ algorithm is first introduced
for spin-1/2 Ising model by Creutz$\left[ 24\right] $ which is a
microcanonical algorithm interpolating between the canonical Monte Carlo and
molecular dynamics techniques on a cellular automaton. The Creutz algorithm
for the spin-1/2 Ising model in two and higher dimensions$\left[ 25,26\right]
$ and spin-1 Ising model in two dimensions$\left[ 27-29\right] $ has been
proven to be successful in producing the value of the static critical
exponents. \ \ \ \ \ \ \ \ 

In the present paper, we investigate the universality of the three
dimensional Blume-Capel model in the second order phase transition region
using the standard and cooling algorithms. For this purpose, the static \
critical exponents are estimated by analyzing \ the data within the frame
work of finite size scaling theory. At the same time, their values are
calculated using power law relations of related thermodynamic quantities \ \
\ Three dimensional Blume-Capel model is expected to be in the universality
class of the three dimensional Ising model for second order phase transition
region with critical exponents $\alpha =0.12$, $\beta =0.31$, $\gamma =1.25$
and $\nu =0.64$ $\left[ 30\right] $. The simulations carried out on simple
cubic lattice $L$x$L$x$L$ of linear dimension $L$ $=8,12,16,20$ and $24$
with periodic boundary conditions. The remainder of the paper is organized
as follows. The details of the model are given in Section 2, the data are
analyzed and the results are discussed in section 3 and a conclusion is
given in Section 4.

\textbf{2 . Model}

Three variables are associated with each site of the lattice. The value of
each site is determined from its value and those of its nearest- neighbors
at the previous time step. The updating rule, which defines a deterministic
cellular automaton, is as follows: Of the three variables on each site, the
first one is the Ising spin $B_{i}$. Its value may be 0 or 1 or 2 . The
Ising spin energy for the model is given by Eq.1. In Eq.1, $S_{i}=B_{i}-1$.
The second variable is for the momentum variable conjugate to the spin ( the
demon ). The kinetic energy associated with the demon, $H_{K}$, is an
integer, which equal to the change in the Ising spin energy for the any spin
flip and its values lie in the interval ($0,$ $m$). The upper limit of the
interval, $m$, is equal to $24J$. The total energy 
\begin{equation}
H=H_{I}+H_{K}
\end{equation}%
is conserved.

The third variable provides a checkerboard style updating, and so it allows
the simulation of the Ising model on a cellular automaton. The black sites
of the checkerboard are updated and then their colour is changed into white;
white sites are changed into black without being updated. The updating rules
for the spin and the momentum variables are as follows: For a site to be
updated its spin is changed one of the other two states with 1/2 probability
and the change in the Ising spin energy ,$dH_{I}$, is calculated. If this
energy change is transferable to or from the momentum variable associated
with this site, such that the total energy $H$ is conserved, then this
change is done and the momentum is appropriately changed. Otherwise the spin
and the momentum are not changed.

For a given total energy the system temperature is obtained from the average
value of kinetic energy, which is given by:\bigskip 
\begin{equation}
\langle E\rangle =\frac{\sum_{n=0}^{m}ne^{-nJ/kT}}{\sum_{n=0}^{m}e^{-nJ/kT}}
\end{equation}
where $E=H_{K}$. The expectation value in Eq. 3 is average over the lattice
and the number of time steps. Because of the third variable, the algorithm
requires two time steps to give every spin of the lattice a chance to
change. Thus, in comparison to ordinary Monte Carlo simulations, two steps
correspond to one full sweep over the system variables.

The initial configurations with the different total energy are obtained
using standard and cooling algorithms$[23]$. In the standard algorithm, all
the spins take the ferromagnetic ordered structure ($\uparrow \uparrow $)
and the kinetic energy is, randomly, given to the lattice via the second
variables in the black sites such that the kinetic energy is equal to the
change in the Ising spin energy for flipped of the spin at those sites. This
initialization procedure resets the starting configuration at each total
energy. The cooling algorithm is divided into two basic parts, the
initialization procedure and the taking of measurements. In the
initialization procedure, firstly, all the spins in the lattice sites take
the ferromagnetic ordered structure ($\uparrow \uparrow $) and the kinetic
energy per site in the all lattice sites is equal to the maximum change in
the Ising spin energy for the any spin flip using the second variables. This
configuration is run during the 10.000 cellular automaton time steps. In the
next step, last configuration in the disordered structure has chosen as a
starting configuration for the cooling run. Rather than resetting the
starting configuration at each energy, it was convenient to use the final
configuration at a given energy as the starting point for the next. During
the cooling cycle, energy was subtracted from the spin system through the
second variables ($H_{k}$) after the 1.000.000 cellular automaton steps.

\textbf{3. Results and discussion}

The universality of the three dimensional Blume-Capel model is investigated
in the continuous transition region using standard and cooling algorithms.
At the cooling algorithm, the cooling rate is equal to 0.01$H_{k}$ per site
for all the $D/J$ values, but \ the kinetic energy of the system reduced by
the different cooling amounts per site because the kinetic energy, $H_{k}$,
is an integer variable in the interval (0, m). The computed values of the
quantities are averages over the lattice and over the number of time steps
(1.000.000) with discard of the first 100.000 time steps during which the
cellular automaton develops.

The critical temperatures are estimated using two different ways. Firstly,
the critical temperatures are estimated from the temperature variation of
the Binder forth-order cumulant$[31,32]$ for the finite lattices. The
Binder\ fourth-order cumulant\ of the magnetization is given by,

\begin{equation}
g_{L}=1-\langle M^{4}\rangle /3\langle M^{2}\rangle ^{2}\quad .
\end{equation}

The temperatures variations of the Binder cumulant are illustrated in
Fig.1(a) for the different lattice sizes at a selected $D/J$ value. The
infinite lattice critical temperatures $T_{c}(\infty ,D/J)$ \ are obtain
from the intersection of the Binder cumulants curves for the different
lattice sizes in Fig1(a). Furthermore, the critical exponent $\nu $ can be
obtained using the finite size scaling relation for the Binder cumulant,
which is defined by$[32]$,

\begin{equation}
g_{L}=G(\varepsilon L^{-1/\nu })
\end{equation}%
where $\varepsilon =(T-T_{c}(\infty ))/T_{c}(\infty )$. It can be seen from
Fig.1(b) that the scaling data for the finite size lattices lie on a single
curve near the critical temperature when the value of the correlation length
critical exponent is equal to the universal value of $\ \nu =0.64$ in the $%
D/J<3$ and $D/J<2.8$ parameter regions at the standard and cooling
algorithm, respectively. \ The scaling of the Binder cumulant for all second
order phase regions exhibits a similar behavior except $D/J=2.8$ value for
cooling algorithm. The scaling of the Binder cumulant at $D/J=2.8$ parameter
value is shown in Fig.2 for standard and cooling algorithm. Although, the
scaling data for finite size lattices lie on single curve with $\nu =0.64$
for standard algorithm(Fig2(a)). The data of Binder cumulant could not be
scaled with $\nu =0.64$ universal value (Fig.2(b)). On the other hand, it is
scaled with $\nu =0.87$ (Fig.2(c)). This result indicates that the value of
the correlation length critical exponent at $D/J=2.8$ is not equal the
theoretical value ($\nu =0.64$) at the cooling algorithm.

Secondly, the critical temperatures are also obtained from susceptibility
maxima $T_{c}^{\chi }(L,D/J)$ and specific heat maxima $T_{c}^{C}(L,D/J)$.
According to finite size scaling theory, the infinite lattice critical
temperature $T_{c}(\infty ,D/J)$ is given by,

\begin{equation}
T_{c}(\infty ,D/J)=T_{c}(L,D/J)+aL^{-1/\nu }.
\end{equation}

For the all $D/J$ values, infinite lattice critical temperatures are
obtained from the extrapolation of susceptibility and specific heat peak
temperatures to$1/L^{1/\nu }$ $\rightarrow 0$ and from intersection of the
Binder cumulant curves at finite lattices. The critical temperature values
estimated from these methods are given in Table 1. The obtained values from
extrapolation and Binder cumulant are in good agreement with each other. On
the other hand, the obtained critical temperatures values at the cooling and
the standard algorithms are \ not in agreement with each other above the $%
D/J=2.2$ value as the increasing of the $D/J$ value. However, the estimated
critical temperatures using the cooling algorithm are in good agreement with
other estimations$[6,8,10,12,16]$ for all $D/J$ parameter values.

The values of the static critical exponents are estimated using power-law
relations and the finite-size scaling relations of thermodynamic quantities.
Firstly, the critical exponents $\beta (L),$ $\gamma (L)$ and $\alpha (L)$
for each lattice are obtained from the log-log plots of the following
power-law relations:\ \ \ \ \ \ \ \ \ 
\begin{eqnarray}
M &=&\varepsilon ^{\beta }\text{ \ \ \ \ \ \ \ \ \ \ \ \ \ \ \ \ \ }%
\varepsilon \rightarrow 0^{-}\text{\ \ } \\
M &=&(-\varepsilon )^{\beta ^{\prime }}\text{ \ \ \ \ \ \ \ \ \ \ \ }%
\varepsilon \rightarrow 0^{+}\text{\ }
\end{eqnarray}%
\begin{eqnarray}
kT\chi &=&\varepsilon ^{-\gamma }\text{ \ \ \ \ \ \ \ \ \ \ \ \ \ }%
\varepsilon \rightarrow 0^{-} \\
kT\chi &=&(-\varepsilon )^{-\gamma ^{\prime }}\text{ \ \ \ \ \ \ \ }%
\varepsilon \rightarrow 0^{+}\text{\ }
\end{eqnarray}%
\begin{eqnarray}
C &=&\varepsilon ^{-\alpha }+b^{-}\text{ \ \ \ \ \ \ \ \ \ }\varepsilon
\rightarrow 0^{-}\text{\ } \\
C &=&(-\varepsilon )^{-\alpha ^{\prime }}+b^{+}\text{ \ \ \ }\varepsilon
\rightarrow 0^{+}
\end{eqnarray}%
where $\varepsilon =(T-T_{c}(L))/T_{c}(L).$ The critical exponents for each
lattice are computed, and these critical exponents are plotted against $%
L^{-1/\nu }$for $T<Tc(\infty ,D/J)$ and $T>Tc(\infty ,D/J)$. The data lie on
straight lines, and their extrapolations to $1/L^{1/\nu }\rightarrow 0$ give
the infinite lattice critical exponents. The estimated infinite lattice
critical exponents($\beta $, $\beta ^{\prime }$, $\gamma $, $\gamma ^{\prime
}$, $\alpha $, $\alpha ^{\prime }$ ) are shown in Table 2, 3 and 4 for all $%
D/J$ values. In the $D/J<3$ parameter region at standard algorithm and in
the $D/J<2.8$ parameter region for cooling algorithm, these values are in
good agreement with theoretical ones. However, the static critical exponents
for $D/J=2.8$ are estimated as $\beta =0.31$, $\gamma =1.58$, $\alpha =0.38$
for $T<Tc(\infty )$ and $\gamma ^{\prime }=1.56$, $\alpha ^{\prime }=0.32$
for $T>Tc(\infty )$ at the cooling algorithm.

The finite size scaling relations of the order parameter M, the
susceptibility $\chi $ and the specific heat C are given by,

\begin{equation}
M=L^{-\beta /\nu }X(\varepsilon L^{-1/\nu })
\end{equation}

\begin{equation}
kT\chi =L^{-\gamma /\nu }Y(\varepsilon L^{-1/\nu })
\end{equation}

\begin{equation}
C=L^{-\alpha /\nu }Z(\varepsilon L^{-1/\nu })
\end{equation}

For large $x=\varepsilon L^{-1/\nu }$, \ the infinite lattice critical
behaviors must be asymptotically reproduced, that is,

\begin{equation}
X(x)=Bx^{\beta }
\end{equation}

\begin{equation}
Y(x)=Cx^{-\gamma }
\end{equation}

\begin{equation}
Z(x)=Ax^{-\alpha }
\end{equation}

The finite size scaling plots of the data for the order parameter M are
shown in Fig.3 for selected $D/J$ values. For $\beta =0.31$ and $\nu =0.64$
theoretical values, the data lie on a single curve for the temperatures both
below and above $T_{c}(\infty ,D/J)$, and validate the finite size scaling
theory in the $D/J<3$ region at standard algorithm and in the $D/J<2.8$
region at cooling algorithm. Thus, the data for M are in agreement with the
universal value of $\beta =0.31$ for $T<T_{c}(\infty ,D/J)$. Also, the
straight line passing through the data for $T>T_{c}(\infty ,D/J)$ behaves
according to Eq. 16 with $\beta ^{\prime }=0.55$ (Fig3(a,b,c)). The order
parameter data for $D/J=2.8$ value are not compatible with the asymptotic
form for $\beta =0.31$ and $\nu =0.64$, but the data lie on a single curve
for $\beta =0.31$ and $\nu =0.87$ for the temperatures below $T_{c}(\infty
,D/J)$ on cooling algorithm. However, the order parameter data do not scale
at this parameter value for the temperatures above $T_{c}(\infty ,D/J)$
(Fig(3(d)).

The finite size scaling plots of the susceptibility are shown in Fig.4 for
the selected $D/J$ values together with the straight lines describing the
theoretically predicted behavior for large x. The scaling of the magnetic
susceptibility data agrees with the asymptotic form with the critical
exponents $\gamma =\gamma ^{\prime }=1.25$ and $\nu =0.64$ for both $%
T<T_{c}(\infty ,D/J)$ and $T>T_{c}(\infty ,D/J)$ in the $D/J<3$ region at
the standard algorithm and $D/J<2.8$ region\ at the cooling algorithm
(Fig.4(a, b, c)). The susceptibility data for $D/J=2.8$ value are not
compatible with the asymptotic form for $\gamma =1.25,$ but the data lies on
a single curve for $\gamma =\gamma ^{\prime }=1.6$ and $\nu =0.87$ for $%
T<T_{c}(\infty )$ and $T>T_{c}(\infty )$ (Fig.4(d)).

The specific heat of an infinite lattice for the Ising model is well
described by$\left[ 33\right] $,

\begin{equation}
C/k=A\varepsilon ^{-a}+b^{\pm }
\end{equation}%
where $b^{\pm }$ express the nonsingular part of the specific heat. The
finite size scaling plots of the singular portion of the specific heat $%
(C/k-b^{\pm })$ are shown in Fig.5 and Fig.6 for the selected $D/J$ values.
The scaling data of specific heat lies on a single curve with the universal
value of \ $\alpha =\alpha ^{\prime }=0.12$ for the temperatures both below
and above $T_{c}(\infty ,D/J)$ in the $D/J<3$ region at the standard
algorithm and in the $D/J<2.8$ region at the cooling algorithm. The data of
the specific heat obtained using cooling algorithm for $D/J=2.8$ scales with 
$\alpha =\alpha ^{\prime }=0.32$ and $\nu =0.87$ at $T<T_{c}(\infty ,D/J)$
and $T>T_{c}(\infty ,D/J).$ The values of $b^{\pm }$ for $T<T_{c}(\infty
,D/J)$ and $T>T_{c}(\infty ,D/J)$ are given in Table4. For the order
parameter, the susceptibility and the specific heat, estimated critical
exponents using finite size scaling theory are in good agreement with
estimated values using power law relations.

To get another estimation for these critical exponents, the finite size
scaling relations at $T=T_{c}(\infty )$ are used. The finite size scaling
relations of the order parameter and the susceptibility at $T_{c}$ are given
by,

\begin{equation}
M=L^{-\beta /\nu }
\end{equation}

\begin{equation}
kT\chi =L^{\gamma /\nu }
\end{equation}

The value of the order parameter and the magnetic susceptibility at $%
T_{c}(\infty ,D/J)$ are determined and the slope obtained from the log-log
plot of the scaling relation corresponding to these quantities gives $\beta
/\nu $ and $\gamma /\nu .$ The estimated values are given in Table 5. The
values of $\beta $ and $\gamma $ obtained from $\beta /\nu $ and $\gamma
/\nu $ using $\nu =0.64$ are in agreement with theoretical values in the $%
D/J<3$ region at standard algorithm and in the $D/J<2.8$ region at the
cooling algorithm. Furthermore, for $D/J=2.8$ parameter value, the value of $%
\beta $ obtained from $\beta /\nu $ is $0.31$ using $\nu =0.87$, and $\gamma 
$ obtained from $\gamma /\nu $ using $\nu =0.87$ is equal to $1.36$ at the
cooling algorithm.

\textbf{4. Conclusion}

The three dimensional Blume-Capel model is simulated using \ two different
procedure which called the standard and cooling algorithms on a cellular
automaton. To determine the universality of the Blume-Capel model which has
a tricritical point at $D/J=2.82$ at the cooling algorithm, \ the static
critical exponents ($\alpha ,\beta ,\gamma $) are estimated using power-law
relations and finite-size scaling relations of the related thermodynamic
quantities. Furthermore, the value of correlation length critical exponent $%
\nu $ is obtained using the finite-size scaling relations of the Binder
cumulant. At the standard algorithm, the model exhibits a continuous phase
transition which compatible with the universal Ising critical behavior in
the $D/J<3$ parameter region. The estimated values of the static critical
exponents( $\beta =0.31$, $\beta ^{\prime }=0.55$, $\gamma =\gamma ^{\prime
}=1.25$, $\alpha =\alpha ^{\prime }=0.12$ and $\nu =0.64$) are independent
on single-ion anisotropy parameter. On the other hand, The cooling algorithm
calculations show that the BC model is compatible with universal Ising
critical behavior in only the $D/J<2.8$ parameter region. Although, the
phase transition is continuous for $D/J=2.8$ value near the tricritical
point($D/J=2.82$).The static critical exponents are not equal to universal
values for second order phase transition. Also, the estimated values ( $%
\beta =0.31$, $\gamma =\gamma ^{\prime }=1.6$, $\alpha =\alpha ^{\prime
}=0.32$ and $\nu =0.87$) of static critical exponents for $D/J=2.8$ are
different from tricritical point critical exponents$[10]$ ($\beta =0.25$, $%
\gamma =1$, $\alpha =0.5$ and $\nu =0.5$). This case indicates that the BC
model exhibit an ununiversal critical behavior at the $D/J=2.8$ parameter
value..

At the same time, the calculated critical temperature values at the cooling
and the standard algorithms are not in agreement with each other above the $%
D/J=2.2$ value as the increasing of the $D/J$ value. However, the estimated
critical temperatures using the cooling algorithm are in good agreement with
other estimations$[10,33]$ for all $D/J$ parameter values. The thermodynamic
quantities at the standard and cooling algorithms behave according to
power-law relations with the different critical exponents for $D/J=2.8$. In
addition, the phase transition occurs at different critical temperature(T$%
_{c}^{g_{L}}=$2.382$\pm 0.001$ at standard algorithm and T$_{c}^{g_{L}}=$1.61%
$\pm 0.02$ at cooling algorithm). The expected behavior for the BC model has
not arisen at the standard algorithm since it does not produced metastable
states in the first order phase transition region. The calculations indicate
that the cooling algorithm is a suitable procedure for the calculations near
the tricritical point, and the BC model exhibits an ununiversal critical
behavior at the $D/J=2.8$ parameter value.

\textbf{Acknowledgements}

This work is supported by a grant from Gazi University (BAP:05/2003-07).

\textbf{References}

[1] M.B. Blume, Phys. Rev. B 141 (1966) 517.

[2] H.W. Capel, Physica(Utrecht) 32 (1966) 966.

[3] W.B. Yelon, D.E.Cox, P.J. Kortman and W.B. Daniels, Phys. Rev. B 9
(1974) 4843.

[4] M. Wortis, Phys. Lett.A 47 (1974) 445.

[5] S.A. Safran, Phys. Rev. Lett. 44 (1980) 937.

[6] A.F. Siqueira and I.P. Fittipaldi, Physica A 138 (1986) 592.

[7] J.W. Tucker, J.Phys.:Condens. Matter I, (1989) 485.

[8] A. Du, H.J. Liu and Y.Q. Y\"{u}, Phys.Stat. Sol. B 241 (2004) 175.

[9] A. Du, Y.Q. Y\"{u} and H.J. Liu, Physica A 320 (2003) 387.

[10] D.M. Saul, M. Wortis and D.Stauffer, Phys.Rev. B 9(1974) 4964.

[11] J.G.Brankov, J.Przystawa and E. Pravecki, J.Phys.C 5 (1972) 3384.

[12] S. Grollau, E. Kierlik, M.L. Rosinberg and G. Tarjus, Phys. Rev. E 63
(2001) 041111-1.

[13] S. Grollau, \ Phys. Rev. E 65 (2002) 056130-1.

[14] A. Falicov and A.N. Berker, Phys. Rev. Lett. 76 (1996) 4380.

[15] G.D. Mahan and S.M. Girvin, Phys.Rev.B 17 (1978) 4411.

[16] C. Buzano and A. Pelizzola, Physica A 216 (1995) 158.

[17] W.M. Ng and J.H.Barry, Phys. Rev. B 17 (1978) 3675.

[18] M. Deserno, Phys. Rev. E. 56 (1997) 5204.

[19] W.G. Wilson and C. A. Vause, Phys. Rew. B 36 (1987) 587.

[20] I. Puha and H.T. Diep, J. Magn. Magn. Mater. 224 (2001) 85.

[21] S.B. Ota and S. Ota, J. Phys.:Condens. Matter I2 (2000) 2233.

[22] S. Ota and S.B. Ota, Phys. Lett. A 285 (2001) 247.

[23] B.Kutlu, A. \"{O}zkan, N. Sefero\u{g}lu, A. Solak and B. Binal, Int. J.
Mod. Phys.C 16 ( 2005) (in press).

[24] M. Creutz , Ann. Phys. 167 (1986) 62; B. Kutlu, S. Turan and M. Kasap,
Int.J.Mod.Phys. C 11 (2000) 561.

[25] B. Kutlu and N.Aktekin, J.Stat. Phys.75 (1994) 757.

[26] N.Aktekin, Physica A 3 436(1995); N.Aktekin, A.G\"{u}nen and Z. Sa\u{g}%
lam, Int.J.Mod.Phys. C 10 (1999) 875.

[27] B.Kutlu, Int.J.Mod.Phys. C 14 (2003) 1305.

[28] A. Solak and B.Kutlu, Int. J. Mod. Phys. C 15 (2004) 1425.

[29] B.Kutlu, Int. J. Mod. Phys.C 12 ( 2001) 1401.

[30] K. Huang, Statistical Mechanics, John Wiley\& Sons, New York, 1987.

[31] K. Binder, Z. Phys. B. Condensed Matter 43 (1981) 119.

[32]V. Privman, Finite Size Scaling and Numerical Simulation of Statistical
Systems, World Scientific, Singapore, 1990.

[33] M.Y.Sykes, D. L.Hunter,D.S.Mckenzie and B.R. Heap, J.Phys.A5 (1972) 667.

[34] S. Saliho\u{g}lu, Tr. J. of Phys., 22 (1998) 1077.

\textbf{Figure Captions}

Fig.1. For the cooling algorithm, (a) The Binder cumulant as a function of $%
kT/J$, (b) finite size scaling plots of the Binder cumulant for $\nu =0.64$.

Fig.2. Finite size scaling plots of the Binder cumulant at $D/J=2.8$ (a) $%
\nu =0.64$ for standard algorithm, (b) $\nu =0.64$ for cooling algorithm,
(c) $\nu =0.87$ for cooling algorithm.

Fig.3. Finite size scaling plots of the order parameter, (a) at $D/J=2.9$
for standard algorithm, (b) $D/J=1$ for cooling algorithm, (c) $D/J=2.8$ for
standard algorithm and (d) $D/J=1$ for cooling algorithm.

Fig.4. Finite size scaling plots of the susceptibility ($\varepsilon
=(T-T_{c}(\infty ))/T_{c}(\infty )$ for $T<T_{c}$, $\varepsilon ^{\prime
}=(T-T_{c}(\infty ))/T)$ for $T>T_{c}$ ), (a) at $D/J=2.8$ for standard
algorithm, (b) $D/J=2.85$ for standard algorithm, (c) $D/J=1$ for cooling
algorithm and (d) $D/J=2.8$ for cooling algorithm, $\varepsilon ^{\prime }$
for $T<T_{c}$ and $T>T_{c}$.

Fig.5. Finite size scaling plots of the specific heat at $T<Tc(\infty ,D/J)$%
, (a) at $D/J=2.8$ for standard algorithm, (b) $D/J=2.8$ for cooling
algorithm, (c) $D/J=2.9$ for standard algorithm and (d) $D/J=1$ for cooling
algorithm.

Fig.6. Finite size scaling plots of the specific heat at $T>Tc(\infty ,D/J)$%
, (a) at $D/J=2.8$ for standard algorithm, (b) $D/J=2.8$ for cooling
algorithm, (c) $D/J=2.9$ for standard algorithm and (d) $D/J=1$ for cooling
algorithm.

\bigskip

\textbf{Table Captions}

Table 1. The estimated infinite lattice critical temperatures $T_{c}(\infty
,D/J)$ for single-ion anisotropy parameter values $(D/J)$.

Table 2. The estimated infinite lattice critical exponents of the order
parameter. $\beta (\infty )$ values are obtained from extrapolation of $%
\beta (L)$ to $1/L^{1/\nu }\rightarrow 0$ .

Table 3. The estimated infinite lattice critical exponents of the
susceptibility. $\gamma (\infty )$ values are obtained from extrapolation of 
$\gamma (L)$ to $1/L^{1/\nu }\rightarrow 0$ .

Table 4. The estimated infinite lattice critical exponents of the specific
heat. $\alpha (\infty )$ values are obtained from extrapolation of $\alpha
(L)$ to $1/L^{1/\nu }\rightarrow 0$ . $b^{\pm }$ values are obtained from
finite size scaling theory.

Table 5. The ${\small \beta /\nu }$ and ${\small \gamma /\nu }$ exponents
are obtained from finite size scaling relations at $T_{c}(\infty ,D/J)$ .

\bigskip 
\newpage%

Table1.\ \ \ \ \ \ \ \ \ \ \ \ \ \ \ \ \ \ \ \ \ \ \ \ \ \ \ \ \ 

$%
\begin{tabular}{|l|c|c|c|c|c|c|}
\hline
& \multicolumn{3}{|c|}{\small standard algorithm} & \multicolumn{3}{|c|}%
{\small cooling algorithm} \\ \hline
\multicolumn{1}{|c|}{${\small D/J}$} & {\small T}$_{c}^{g_{L}}(\infty )$ & 
{\small T}$_{c}^{C}(\infty )$ & {\small T}$_{c}^{\chi }(\infty )$ & {\small T%
}$_{c}^{g_{L}}(\infty )$ & {\small T}$_{c}^{C}(\infty )$ & {\small T}$%
_{c}^{\chi }(\infty )$ \\ \hline
\multicolumn{1}{|c|}{${\small -2}$} & {\small 3.63}$\pm 0.01$ & {\small 3.62}%
$\pm 0.01$ & {\small 3.60}$\pm 0.02$ & {\small 3.63}$\pm 0.01$ & {\small 3.63%
}$\pm 0.02$ & {\small 3.64}$\pm 0.02$ \\ \hline
\multicolumn{1}{|c|}{${\small 0}$} & {\small 3.20}$\pm 0.01$ & {\small 3.20}$%
\pm 0.02$ & {\small 3.19}$\pm 0.02$ & {\small 3.20}$\pm 0.01$ & {\small 3.20}%
$\pm 0.02$ & {\small 3.20}$\pm 0.01$ \\ \hline
\multicolumn{1}{|c|}{${\small 1}$} & {\small 2.88}$\pm 0.01$ & {\small 2.88}$%
\pm 0.01$ & {\small 2.88}$\pm 0.02$ & {\small 2.88}$\pm 0.01$ & {\small 2.88}%
$\pm 0.03$ & {\small 2.87}$\pm 0.01$ \\ \hline
\multicolumn{1}{|c|}{${\small 2}$} & {\small 2.41}$\pm 0.01$ & {\small 2.40}$%
\pm 0.01$ & {\small 2.39}$\pm 0.02$ & {\small 2.42}$\pm 0.01$ & {\small 2.41}%
$\pm 0.02$ & {\small 2.43}$\pm 0.01$ \\ \hline
\multicolumn{1}{|c|}{${\small 2.2}$} & {\small 2.13}$\pm 0.01$ & {\small 2.13%
}$\pm 0.03$ & {\small 2.13}$\pm 0.03$ & {\small 2.27}$\pm 0.02$ & {\small %
2.28}$\pm 0.02$ & {\small 2.27}$\pm 0.01$ \\ \hline
\multicolumn{1}{|c|}{${\small 2.4}$} & {\small 2.17}$\pm 0.01$ & {\small 2.17%
}$\pm 0.03$ & {\small 2.17}$\pm 0.03$ & {\small 2.11}$\pm 0.02$ & {\small %
2.10}$\pm 0.02$ & {\small 2.12}$\pm 0.02$ \\ \hline
\multicolumn{1}{|c|}{${\small 2.6}$} & {\small 2.40}$\pm 0.01$ & {\small 2.42%
}$\pm 0.02$ & {\small 2.40}$\pm 0.02$ & {\small 1.93}$\pm 0.03$ & {\small %
1.95}$\pm 0.03$ & {\small 1.93}$\pm 0.05$ \\ \hline
\multicolumn{1}{|c|}{${\small 2.8}$} & {\small 2.38}$\pm 0.01$ & {\small 2.39%
}$\pm 0.01$ & {\small 2.38}$\pm 0.02$ & {\small 1.61}$\pm 0.05$ & {\small %
1.60}$\pm 0.02$ & {\small 1.61}$\pm 0.02$ \\ \hline
\multicolumn{1}{|c|}{${\small 2.82}$} & {\small 2.31}$\pm 0.01$ & {\small %
2.30}$\pm 0.02$ & {\small 2.31}$\pm 0.02$ & {\small 1.59}$\pm 0.03$ & 
{\small 1.61}$\pm 0.03$ & {\small 1.60}$\pm 0.04$ \\ \hline
\multicolumn{1}{|c|}{${\small 2.85}$} & {\small 2.36}$\pm 0.01$ & {\small %
2.36}$\pm 0.02$ & {\small 2.36}$\pm 0.02$ & {\small 1.39}$\pm 0.01$ & 
{\small 1.41}$\pm 0.02$ & {\small 1.41}$\pm 0.02$ \\ \hline
\multicolumn{1}{|c|}{${\small 2.9}$} & {\small 2.38}$\pm 0.01$ & {\small 2.37%
}$\pm 0.03$ & {\small 2.39}$\pm 0.03$ & {\small 1.30}$\pm 0.01$ & {\small %
1.29}$\pm 0.02$ & {\small 1.28}$\pm 0.02$ \\ \hline
\end{tabular}%
$

\bigskip

Table2.

\begin{tabular}{|c|c|c|c|c|}
\hline
& \multicolumn{2}{|c|}{\small standard algorithm} & \multicolumn{2}{|c|}%
{\small cooling algorithm} \\ \hline
${\small D/J}$ & ${\small \beta }$ & ${\small \beta }^{\prime }$ & ${\small %
\beta }$ & ${\small \beta }^{\prime }$ \\ \hline
${\small -2}$ & {\small 0.29}$\pm 0.02$ & {\small 0.58}$\pm 0.01$ & {\small %
0.30}$\pm 0.03$ & {\small 0.54}$\pm 0.02$ \\ \hline
${\small 0}$ & {\small 0.29}$\pm 0.06$ & {\small 0.57}$\pm 0.02$ & {\small %
0.31}$\pm 0.02$ & {\small 0.57}$\pm 0.01$ \\ \hline
${\small 1}$ & {\small 0.30}$\pm 0.06$ & {\small 0.55}$\pm 0.01$ & {\small %
0.31}$\pm 0.02$ & {\small 0.54}$\pm 0.01$ \\ \hline
${\small 2}$ & {\small 0.30}$\pm 0.01$ & {\small 0.57}$\pm 0.01$ & {\small %
0.31}$\pm 0.01$ & {\small 0.52}$\pm 0.01$ \\ \hline
${\small 2.2}$ & {\small 0.31}$\pm 0.02$ & {\small 0.57}$\pm 0.06$ & {\small %
0.31}$\pm 0.03$ & {\small 0.51}$\pm 0.03$ \\ \hline
${\small 2.4}$ & {\small 0.32}$\pm 0.01$ & {\small 0.57}$\pm 0.01$ & {\small %
0.31}$\pm 0.02$ & {\small 0.51}$\pm 0.04$ \\ \hline
${\small 2.6}$ & {\small 0.32}$\pm 0.06$ & {\small 0.55}$\pm 0.06$ & {\small %
0.31}$\pm 0.03$ & {\small 0.52}$\pm 0.01$ \\ \hline
${\small 2.8}$ & {\small 0.33}$\pm 0.01$ & {\small 0.56}$\pm 0.02$ & {\small %
0.31}$\pm 0.02$ & {\small -} \\ \hline
${\small 2.82}$ & {\small 0.33}$\pm 0.06$ & {\small 0.57}$\pm 0.06$ & 
{\small -} & {\small -} \\ \hline
${\small 2.85}$ & {\small 0.33}$\pm 0.01$ & {\small 0.56}$\pm 0.01$ & 
{\small -} & {\small -} \\ \hline
${\small 2.9}$ & {\small 0.34}$\pm 0.01$ & {\small 0.58}$\pm 0.01$ & {\small %
-} & {\small -} \\ \hline
\end{tabular}
\ \ \ \ \ 

\bigskip

\pagebreak%

Table 3.\ \ \ 

\bigskip 
\begin{tabular}{|c|c|c|c|c|}
\hline
& \multicolumn{2}{|c|}{\small standard algorithm} & \multicolumn{2}{|c|}%
{\small cooling algorithm} \\ \hline
{\small D/J} & ${\small \gamma }$ & ${\small \gamma }^{\prime }$ & ${\small %
\gamma }$ & ${\small \gamma }^{\prime }$ \\ \hline
{\small -2} & {\small 1.27}$\pm 0.06$ & {\small 1.21}$\pm 0.06$ & {\small %
1.24}$\pm 0.03$ & {\small 1.21}$\pm 0.03$ \\ \hline
{\small 0} & {\small 1.26}$\pm 0.01$ & {\small 1.18}$\pm 0.02$ & {\small 1.24%
}$\pm 0.04$ & {\small 1.22}$\pm 0.02$ \\ \hline
{\small 1} & {\small 1.29}$\pm 0.01$ & {\small 1.15}$\pm 0.02$ & {\small 1.26%
}$\pm 0.06$ & {\small 1.21}$\pm 0.04$ \\ \hline
{\small 2} & {\small 1.26}$\pm 0.01$ & {\small 1.18}$\pm 0.02$ & {\small 1.22%
}$\pm 0.02$ & {\small 1.21}$\pm 0.03$ \\ \hline
{\small 2.2} & {\small 1.26}$\pm 0.01$ & {\small 1.13}$\pm 0.01$ & {\small %
1.26}$\pm 0.04$ & {\small 1.26}$\pm 0.03$ \\ \hline
{\small 2.4} & {\small 1.26}$\pm 0.01$ & {\small 1.13}$\pm 0.01$ & {\small %
1.24}$\pm 0.03$ & {\small 1.24}$\pm 0.02$ \\ \hline
{\small 2.6} & {\small 1.26}$\pm 0.01$ & {\small 1.20}$\pm 0.02$ & {\small %
1.23}$\pm 0.05$ & {\small 1.21}$\pm 0.06$ \\ \hline
{\small 2.8} & {\small 1.25}$\pm 0.02$ & {\small 1.25}$\pm 0.01$ & {\small %
1.58}$\pm 0.04$ & {\small 1.56}$\pm 0.05$ \\ \hline
{\small 2.82} & {\small 1.19}$\pm 0.06$ & {\small 1.25}$\pm 0.01$ & {\small -%
} & {\small -} \\ \hline
{\small 2.85} & {\small 1.25}$\pm 0.03$ & {\small 1.12}$\pm 0.02$ & {\small -%
} & {\small -} \\ \hline
{\small 2.9} & {\small 1.27}$\pm 0.06$ & {\small 1.16}$\pm 0.02$ & {\small -}
& {\small -} \\ \hline
\end{tabular}
\ \ \ \ \ \ \ \ \ \ \ \ \ \ \ 

\bigskip

Table 4

\begin{tabular}{|c|c|c|c|c|c|c|c|c|}
\hline
& \multicolumn{4}{|c|}{\small standard algorithm} & \multicolumn{4}{|c|}%
{\small cooling algorithm} \\ \hline
{\small D/J} & ${\small \alpha }$ & ${\small \alpha }^{\prime }$ & ${\small b%
}^{-}$ & ${\small b}^{+}$ & ${\small \alpha }$ & ${\small \alpha }^{\prime }$
& ${\small b}^{-}$ & ${\small b}^{+}$ \\ \hline
{\small -2} & {\small 0.13}$\pm 0.03$ & {\small 0.12}$\pm 0.01$ & {\small %
-0.0005} & {\small -0.025} & {\small 0.12}$\pm 0.02$ & {\small 0.12}$\pm
0.02 $ & {\small -0.005} & {\small -0.028} \\ \hline
{\small 0} & {\small 0.12}$\pm 0.02$ & {\small 0.11}$\pm 0.01$ & {\small %
-0.005} & {\small -0.08} & {\small 0.12}$\pm 0.02$ & {\small 0.12}$\pm 0.03$
& {\small -0.005} & {\small -0.08} \\ \hline
{\small 1} & {\small 0.13}$\pm 0.03$ & {\small 0.12}$\pm 0.03$ & {\small 0}
& {\small -0.54} & {\small 0.13}$\pm 0.03$ & {\small 0.11}$\pm 0.02$ & 
{\small -0.01} & {\small -0.15} \\ \hline
{\small 2} & {\small 0.12}$\pm 0.02$ & {\small 0.12}$\pm 0.03$ & {\small %
-0.007} & {\small -0.4} & {\small 0.13}$\pm 0.05$ & {\small 0.12}$\pm 0.01$
& {\small -0.09} & {\small -0.4} \\ \hline
{\small 2.2} & {\small 0.11}$\pm 0.06$ & {\small 0.12}$\pm 0.03$ & {\small %
-0.006} & {\small -0.81} & {\small 0.13}$\pm 0.05$ & {\small 0.12}$\pm 0.04$
& {\small -0.3} & {\small -0.6} \\ \hline
{\small 2.4} & {\small 0.11}$\pm 0.02$ & {\small 0.12}$\pm 0.02$ & {\small 0}
& {\small -1.35} & {\small 0.12}$\pm 0.03$ & {\small 0.12}$\pm 0.03$ & 
{\small 0} & {\small -0.65} \\ \hline
{\small 2.6} & {\small 0.11}$\pm 0.03$ & {\small 0.11}$\pm 0.01$ & {\small 0}
& {\small -18} & {\small 0.13}$\pm 0.04$ & {\small 0.13}$\pm 0.05$ & {\small %
-0.15} & {\small -1.6} \\ \hline
{\small 2.8} & {\small 0.12}$\pm 0.02$ & {\small 0.10}$\pm 0.01$ & {\small 0}
& {\small -5.5} & {\small 0.38}$\pm 0.07$ & {\small 0.32}$\pm 0.07$ & 
{\small -5} & {\small 0} \\ \hline
{\small 2.82} & {\small 0.12}$\pm 0.02$ & {\small 0.12}$\pm 0.01$ & {\small 0%
} & {\small -7.5} & {\small -} & {\small -} & {\small -} & {\small -} \\ 
\hline
{\small 2.85} & {\small 0.11}$\pm 0.02$ & {\small 0.11}$\pm 0.01$ & {\small %
-0.02} & {\small -12} & {\small -} & {\small -} & {\small -} & {\small -} \\ 
\hline
{\small 2.9} & {\small 0.13}$\pm 0.02$ & {\small 0.11}$\pm 0.01$ & {\small %
-0.06} & {\small -15} & {\small -} & {\small -} & {\small -} & {\small -} \\ 
\hline
\end{tabular}

\bigskip

\newpage%

Table 5.

\begin{tabular}{|c|c|c|c|c|}
\hline
& \multicolumn{2}{|c|}{\small standard algorithm} & \multicolumn{2}{|c|}%
{\small cooling algorithm} \\ \hline
${\small D/J}$ & ${\small \beta /\nu }$ & ${\small \gamma /\nu }$ & ${\small %
\beta /\nu }$ & ${\small \gamma /\nu }$ \\ \hline
${\small -2}$ & {\small 0.49}$\pm 0.04$ & {\small 1.95}$\pm 0.03$ & {\small %
0.48}$\pm 0.06$ & {\small 1.94}$\pm 0.04$ \\ \hline
${\small 0}$ & {\small 0.49}$\pm 0.05$ & {\small 1.95}$\pm 0.03$ & {\small %
0.48}$\pm 0.02$ & {\small 1.93}$\pm 0.04$ \\ \hline
${\small 1}$ & {\small 0.52}$\pm 0.02$ & {\small 1.92}$\pm 0.02$ & {\small %
0.48}$\pm 0.03$ & {\small 1.93}$\pm 0.03$ \\ \hline
${\small 2}$ & {\small 0.44}$\pm 0.07$ & {\small 1.86}$\pm 0.03$ & {\small %
0.47}$\pm 0.03$ & {\small 1.66}$\pm 0.03$ \\ \hline
${\small 2.2}$ & {\small 0.50}$\pm 0.03$ & {\small 1.91}$\pm 0.03$ & {\small %
0.49}$\pm 0.02$ & {\small 1.92}$\pm 0.01$ \\ \hline
${\small 2.4}$ & {\small 0.49}$\pm 0.03$ & {\small 1.95}$\pm 0.02$ & {\small %
0.50}$\pm 0.06$ & {\small 1.71}$\pm 0.02$ \\ \hline
${\small 2.6}$ & {\small 0.48}$\pm 0.03$ & {\small 1.82}$\pm 0.03$ & {\small %
0.48}$\pm 0.04$ & {\small 1.95}$\pm 0.09$ \\ \hline
${\small 2.8}$ & {\small 0.50}$\pm 0.05$ & {\small 1.82}$\pm 0.03$ & {\small %
0.36}$\pm 0.05$ & {\small 1.56}$\pm 0.07$ \\ \hline
${\small 2.82}$ & {\small 0.49}$\pm 0.04$ & {\small 1.89}$\pm 0.01$ & 
{\small -} & {\small -} \\ \hline
${\small 2.85}$ & {\small 0.49}$\pm 0.02$ & {\small 1.87}$\pm 0.02$ & 
{\small -} & {\small -} \\ \hline
${\small 2.9}$ & {\small 0.49}$\pm 0.03$ & {\small 1.84}$\pm 0.03$ & {\small %
-} & {\small -} \\ \hline
\end{tabular}

\bigskip

\bigskip

\bigskip \FRAME{ftbpF}{6.5838in}{3.9634in}{0in}{}{}{fig1ab.eps}{\special%
{language "Scientific Word";type "GRAPHIC";maintain-aspect-ratio
TRUE;display "USEDEF";valid_file "F";width 6.5838in;height 3.9634in;depth
0in;original-width 6.5215in;original-height 3.915in;cropleft "0";croptop
"1";cropright "1";cropbottom "0";filename 'fig1ab.eps';file-properties
"XNPEU";}}

\bigskip

\bigskip \FRAME{ftbpF}{6.4947in}{3.6357in}{0pt}{}{}{fig2abc.eps}{\special%
{language "Scientific Word";type "GRAPHIC";maintain-aspect-ratio
TRUE;display "USEDEF";valid_file "F";width 6.4947in;height 3.6357in;depth
0pt;original-width 6.7014in;original-height 3.7395in;cropleft "0";croptop
"1";cropright "1";cropbottom "0";filename 'fig2abc.eps';file-properties
"XNPEU";}}

\bigskip

\bigskip \FRAME{ftbpF}{6.0675in}{6.3123in}{0in}{}{}{fig3abcd.eps}{\special%
{language "Scientific Word";type "GRAPHIC";maintain-aspect-ratio
TRUE;display "USEDEF";valid_file "F";width 6.0675in;height 6.3123in;depth
0in;original-width 6.0079in;original-height 6.2509in;cropleft "0";croptop
"1";cropright "1";cropbottom "0";filename 'fig3abcd.eps';file-properties
"XNPEU";}}

\bigskip

\FRAME{ftbpF}{6.0226in}{7.0257in}{0pt}{}{}{fig4abcd.eps}{\special{language
"Scientific Word";type "GRAPHIC";display "USEDEF";valid_file "F";width
6.0226in;height 7.0257in;depth 0pt;original-width 5.9196in;original-height
10.7133in;cropleft "0";croptop "0.9286";cropright "1.0385";cropbottom
"0";filename 'fig4abcd.eps';file-properties "XNPEU";}}

\bigskip

\bigskip \FRAME{ftbpF}{6.0226in}{7.0257in}{0pt}{}{}{fig5abcd.eps}{\special%
{language "Scientific Word";type "GRAPHIC";display "USEDEF";valid_file
"F";width 6.0226in;height 7.0257in;depth 0pt;original-width
6.0753in;original-height 10.3319in;cropleft "0";croptop "1.0249";cropright
"1.1663";cropbottom "0";filename 'fig5abcd.eps';file-properties "XNPEU";}}

\bigskip \FRAME{ftbpF}{6.0226in}{7.0257in}{0pt}{}{}{fig6abcd.eps}{\special%
{language "Scientific Word";type "GRAPHIC";display "USEDEF";valid_file
"F";width 6.0226in;height 7.0257in;depth 0pt;original-width
5.9049in;original-height 9.5752in;cropleft "0";croptop "1.0251";cropright
"1.1621";cropbottom "0";filename 'fig6abcd.eps';file-properties "XNPEU";}}

\bigskip

\bigskip

\bigskip

\bigskip

\bigskip

\bigskip

\bigskip

\bigskip

\bigskip

\bigskip

\bigskip

\bigskip

\bigskip

\bigskip

\bigskip

\bigskip

\bigskip

\bigskip

\bigskip

\bigskip

\end{document}